\documentstyle[maggie]{article}

\def\be{\begin{equation}}
\def\ee{\end{equation}}
\def\bea{\begin{eqnarray}}
\def\eea{\end{eqnarray}}
\def\eelc{$e^+e^-$ linear colliders}

\begin{document}
\begin{flushright}
PSI-PR-04-10 \\
hep-ph/0409200 
\end{flushright}
\vspace*{0.3cm}

\title{SDECAY - A Fortran Code For SUSY Particle Decays In The MSSM\\}

\author{M.~M.~M\"UHLLEITNER}

\address{Paul Scherrer Institute, CH-5232 Villigen PSI, Switzerland}

%\author{A.~DJOUADI}

%\address{LPMT, Universit\'e de Montpellier II, F-34095 Montpellier Cedex 5, 
%France \\
%LPTHE, Universit\'es de Paris 6\&7, 4 Place Jussieu, F-75252 Paris
%Cedex 05}

%\author{Y.~MAMBRINI}

%\address{Departamento de F\'{\i}sica Te\'orica C-XI and Instituto de 
%F\'{\i}sica Te\'orica C-XVI, Universidad Aut\'onoma de Madrid, Cantoblanco,
%28049 Madrid, Spain}


\maketitle\abstracts{
%The Fortran code {\tt SDECAY} calculates the decay widths and branching
%ratios of all supersymmetric (SUSY) particles in the Minimal Supersymmetric
%Extension of the Standard Model (MSSM). The two-body decays of sfermions and
%gauginos have been implemented as well as the three-body decays of 
%charginos, neutralinos, the gluino and the top and bottom squarks. Also 
%included are the four-body decays of the stop and the important 
%loop-induced decay modes. In addition, the QCD corrections to the 
%two-body decays involving strongly interacting particles and the dominant
%components of the electroweak corrections to all decay modes are implemented. 
 The Fortran code {\tt SDECAY} calculates the decay widths
 and branching ratios of all supersymmetric (SUSY) particles in the Minimal
 Supersymmetric Standard Model (MSSM), including higher order effects. 
 The usual
 two-body decays of sfermions and gauginos as well as the three-body
 decay modes of charginos, neutralinos and gluinos are included.
 Furthermore, the three-body stop and sbottom decays and even the 
 four-body stop 
 decays are calculated. The important loop-induced decays, the QCD
 corrections to the two-body widths involving strongly interacting
 particles and the dominant electroweak effects to all processes are
 evaluated as well.
}

%***********************************************************************
\section{Introduction}
%***********************************************************************
The search for SUSY particles is a major goal at present and future colliders.
Once they are found, their properties are expected to be determined with an
accuracy of a few per cent at the LHC and at the per cent level and below
at future \eelc. To match the expected experimental accuracy, programs 
are needed for the calculation of the SUSY particle spectrum, of the 
production cross sections and of the SUSY particle decay widths and 
branching ratios with high precision, also including higher order effects.
The Fortran code {\tt SDECAY}\cite{sdecay} which is presented in the 
following deals with the decay of SUSY particles in the MSSM and includes
the most important higher order effects.
%\footnote{The code can be obtained
%at the url: http://people.web.psi.ch/muehlleitner/SDECAY} 
Up to the user's
choice, the mass spectrum and soft SUSY breaking parameters are obtained from 
the renormalization group evolution program {\tt SuSpect}\cite{suspect} or 
from an input file in the SUSY Les Houches Accord (SLHA) format\cite{slha}. 
{\tt SDECAY} then evaluates the various couplings of the SUSY particles and 
MSSM Higgs bosons and calculates the decay widths and branching 
ratios.\footnote{For details of the implementation of the MSSM see the user's
manual\cite{sdecay}.} 

\section{The tree-level and QCD corrected two-body decays}
The following tree-level two-body sfermion decays have been implemented in 
the program\footnote{Here and in the following the distinction between the 
two isospin (s)fermion partners is discarded.}
\bea
\tilde{f}_i &\to& \tilde{\chi}_j f  \qquad 
\tilde{\chi}_j:\;{\mathrm chargino,\;
neutralino}  \nonumber\\
\tilde{f}_i &\to& \tilde{f}_j V \qquad V:\; W,Z \\
\tilde{f}_i &\to& \tilde{f}_j \Phi \qquad \Phi:\; h,H,A,H^\pm 
\nonumber
\eea
The included two-body decays for the squarks, the gluino, the charginos and
neutralinos are 
\bea
\tilde{q}_i &\to& \tilde{g}\, q, \quad\quad
\tilde{g} \to \tilde{q}_i q \\
\tilde{\chi}_i &\to& \tilde{\chi}_j V, \quad\;
\tilde{\chi}_i \to \tilde{\chi}_j \Phi \quad {\mathrm and}\quad
\tilde{\chi}_i \to \tilde{f}_j f 
\eea
In case of the gauge mediated SUSY breaking model, the following 2-body decays
into the lightest SUSY particle, the gravitino, have been implemented\cite{sdecay}
\bea
\tilde{\chi}_1^0 \to \tilde{G}\gamma, \tilde{G}Z, \tilde{G}\phi
\;\;(\phi=h,H,A)\quad {\mathrm and} \quad \tilde{\tau}_1 \to \tilde{G}\tau
\eea
%In the various couplings, the masses of the third-generation fermions are 
%the running $\overline{{\mathrm DR}}$ masses at the scale of the electroweak 
%symmetry 
%breaking (EWSB). This is also the case for all soft-SUSY breaking parameters 
%and the third-generation sfermion mixing angles entering the couplings. In  
%addition, the user has the option to choose the value of the QCD coupling
%constant and the $b,t$ Yukawa couplings evaluated at the scale of the 
%decaying SUSY particle or any other scale. \s

%A new feature of the program {\tt SDECAY} compared to existing programs for
%SUSY particle decays is the calculation of the SUSY QCD corrections to 
%2-body decays involving coloured particles\cite{sdecay}:
A new feature compared to many existing decay programs, is the inclusion of 
the SUSY-QCD corrections to 2-body decays involving coloured 
particles\cite{sdecay}:
\bea
\tilde{q}_i &\to& \tilde{\chi}_j q,\quad \tilde{q}_i \to \tilde{q}_j V,\quad
\tilde{q}_i \to \tilde{q}_j \Phi,\quad \tilde{q}_i \to \tilde{g} q \\
\tilde{g} &\to& \tilde{q}_i q, \quad
\tilde{\chi}_i \to \tilde{q}_j q
\eea
We are taking running parameters at the scale of the electroweak symmetry
breaking for the gauge and third generation Yukawa couplings as well as the
soft SUSY-breaking parameters and third-generation sfermion mixing angles.
In this way, the bulk of the electroweak corrections has been taken into
account.
%The bulk of the electroweak corrections has been accounted for by taking the
%gauge and third generation Yukawa couplings as well as the soft SUSY-breaking 
%parameters and third-generation sfermion mixing angles entering the 
%couplings at the scale of the electroweak symmetry breaking. 
%In addition, the user has the option to choose the value of the QCD coupling
%constant and the $b,t$ Yukawa couplings evaluated at the scale of the 
%decaying SUSY particle or any other scale.
Finally, {\tt SDECAY} also evaluates the top quark decays in the MSSM:
\bea
t \to bW^+, \quad t\to bH^+ \quad{\mathrm and}\quad
t\to \tilde{t}_1 \tilde{\chi}_1^0 
\eea

\section{The loop-induced and the multibody decays}
If the 2-body decays are closed, multibody decays will be dominant. 
The implemented gaugino, gluino, stop and sbottom three-body
decays\cite{sdecay} are
\bea
\tilde{\chi}_i &\to& \tilde{\chi}_j f\bar{f}, \quad\quad
\tilde{\chi}_i \to \tilde{g} q\bar{q} \\
\tilde{g} &\to& \tilde{\chi}_i q\bar{q}, \quad 
\;\quad\tilde{g} \to \tilde{t}_1 \bar{b} W^-, \quad
\tilde{g} \to \tilde{t}_1 \bar{b} H^- \\
\tilde{t}_i &\to& bW^+\tilde{\chi}_1^0, \quad \tilde{t}_i 
\to bH^+\tilde{\chi}_1^0 \\
\tilde{t}_i &\to& bl^+\tilde{\nu}_l, \quad \tilde{t}_i \to
b\tilde{l}^+\nu_l, \quad
\tilde{t}_i \to \tilde{b}_j f \bar{f}, \quad 
\tilde{t}_2 \to \tilde{t}_1 f \bar{f} \\
\tilde{b}_i &\to& tl^-\tilde{\nu}_l^*, \quad\! 
\tilde{b}_i \to t\tilde{l}^- \bar{\nu}_l, \quad
\tilde{b}_i \to \tilde{t}_j f \bar{f}, \quad 
\tilde{b}_2 \to \tilde{b}_1 f\bar{f}
\eea
In the 3-body decays, the mixing of the third-generation sfermions as well as 
the final-state fermion masses have been taken into account\cite{sdecay,revis}.
Moreover, loop-induced decays of the lightest stop, gluino and
the next-to-lightest neutralino\cite{sdecay} might become important and have 
therefore been included:
\bea
\tilde{\chi}_2^0 \to \tilde{\chi}_1^0 \gamma, \qquad
\tilde{g} \to \tilde{\chi}_1^0 g \qquad{\mathrm and}\qquad
\tilde{t}_1 \to \tilde{\chi}_1^0 c
\eea

If the 3-body decays are kinematically forbidden, the 
following $\tilde{t}_1$ 4-body decay can compete with the loop-induced 
$\tilde{t}_1$ decay and has been implemented:
\bea
\tilde{t}_1 &\to& b\tilde{\chi}_1^0 f\bar{f} 
\eea

\section{Structure of {\tt SDECAY}}
Besides the necessary {\tt SuSpect} files, the program 
{\tt SDECAY} is composed of 3 files: (i) {\tt sdecay.in}, where the 
user can choose various options, such as whether or not to include
the QCD corrections to 2-body decays, the multibody and/or loop decays, 
the GMSB decays, the top decays. One can choose the scale and the
number of loops for the running couplings and also, where to take the 
spectrum from, {\it i.e.}~from {\tt SuSpect} or any input file in the
SLHA format; (ii) {\tt sdecay.f} is the main body of the code. It evaluates 
the various couplings and calculates the decay widths and branching ratios; 
(iii) {\tt sdecay.out}, which gives out the branching ratios and total widths,
either in a simple transparent form
or in the SLHA format, which is up to the user's choice. In addition, the 
whole SUSY spectrum, mixing matrices and further parameters used for the 
calculation of the couplings and decay widths are written out. The necessary 
files and the makefile for the compilation can be found at the {\tt SDECAY} 
homepage:
\\
\centerline{http://people.web.psi.ch/muehlleitner/SDECAY/.} 

\section{Summary}
The Fortran code {\tt SDECAY} calculates all SUSY particle 2-body tree-level
and QCD corrected decays, 3-body and loop-decays as well as $\tilde{t}_1$ 
4-body and top-decays. 
%It is at present the only program including higher order corrections in the 
%partial widths. 
The program is rather fast, flexible and maintained regularly 
to include upgrades and newest theoretical developments as well as experimental
needs. 
%Any suggestions from potential users will be welcome.

\end{document}